\documentclass[10pt,aps,prl,reprint,twocolumn,bibnotes,footinbib,amsfonts,amssymb,amsmath,preprintnumbers,notitlepage,superscriptaddress,floats,floatfix]{revtex4-1}
\usepackage[utf8]{inputenc}
\usepackage[british]{babel}
\usepackage[Symbolsmallscale]{upgreek}
\usepackage{isomath}
\usepackage[protrusion=true,tracking=true,kerning=true,spacing=true,final,babel=true]{microtype}
\usepackage{physics}
\usepackage{xcolor}
\usepackage{graphicx}
\usepackage[final]{hyperref}

\begin{document}
\title{Anisotropies in the Astrophysical Gravitational-Wave Background:\\The Impact of Black Hole Distributions}

\author{Alexander~C.~Jenkins}
\email{alexander.jenkins@kcl.ac.uk}
\affiliation{Theoretical Particle Physics and Cosmology Group, Physics Department, King's College London, University of London, Strand, London WC2R 2LS, United Kingdom}

\author{Richard~O`Shaughnessy}
\email{richard.oshaughnessy@ligo.org}
\affiliation{Center for Computational Relativity and Gravitation, Rochester Institute of Technology, 85 Lomb Memorial Drive, Rochester, New York 14623, USA}

\author{Mairi~Sakellariadou}
\email{mairi.sakellariadou@kcl.ac.uk}
\affiliation{Theoretical Particle Physics and Cosmology Group, Physics Department, King's College London, University of London, Strand, London WC2R 2LS, United Kingdom}

\author{Daniel~Wysocki}
\email{dw2081@rit.edu}
\affiliation{Center for Computational Relativity and Gravitation, Rochester Institute of Technology, 85 Lomb Memorial Drive, Rochester, New York 14623, USA}

\date{\today}
\preprint{KCL-PH-TH/2018-60}

\begin{abstract}
    We use population inference to explore the impact that uncertainties in the distribution of binary black holes (BBH) have on the astrophysical gravitational-wave background (AGWB).
    Our results show that the AGWB monopole is sensitive to the nature of the BBH population (particularly the local merger rate), while the anisotropic $C_\ell$ spectrum is only modified to within a few percent, at a level which is insignificant compared to other sources of uncertainty (such as cosmic variance).
    This is very promising news for future observational studies of the AGWB, as it shows that (i) the monopole can be used as a new probe of the population of compact objects throughout cosmic history, complementary to direct observations by LIGO and Virgo and (ii) we are able to make surprisingly robust predictions for the $C_\ell$ spectrum, even with only very approximate knowledge of the black hole population.
    As a result, the AGWB anisotropies have enormous potential as a new probe of the large-scale structure of the Universe, and of late-Universe cosmology in general.
\end{abstract}
\maketitle

%%%%%%%%%%%%%%%%%%%%%%%%%%%%%%%%%%%%%%%%%%%%%%%%%%%%%%%%%%%%%%%%%%%%%%%%%%%%%%%%%
\paragraph{Introduction.}$\!\!\!\!$---
Following the direct detection of gravitational waves (GWs) by the Advanced LIGO~\cite{TheLIGOScientific:2014jea} and Advanced Virgo~\cite{TheVirgo:2014hva} interferometers, we have entered an era of GW astronomy.
Eleven GW detections have been confirmed so far~\cite{LIGOScientific:2018mvr,Abbott:2016blz,Abbott:2016nmj,TheLIGOScientific:2016pea,Abbott:2017vtc,Abbott:2017oio,TheLIGOScientific:2017qsa,Abbott:2017gyy}, each originating from a compact binary coalescence (CBC).
As LIGO and Virgo move towards design sensitivity, and as further advanced interferometers come online~\cite{Aso:2013eba,Audley:2017drz,Punturo:2010zz}, we expect to observe ever-increasing numbers of CBCs~\cite{Aasi:2013wya}, giving us unprecedented knowledge about the population of black holes (BHs) and neutron stars (NSs) in the local Universe.

However, many CBCs will remain unresolvable, particularly those at higher redshift.
The superposition of many coincident CBCs (resolvable and unresolvable) leads to the astrophysical gravitational-wave background (AGWB), a persistent GW signal that can be distinguished from the instrumental noise by correlating data from multiple detectors~\cite{Romano:2016dpx,Smith:2017vfk}.
The AGWB is predicted to be detected soon after LIGO and Virgo reach design sensitivity in 2022, or if optimistic forecasts are borne out, possibly even during the next observing run in 2019 (based on the CBC rates and mass distributions given in Refs.~\cite{Abbott:2017vtc,TheLIGOScientific:2017qsa}, as well as the future detector sensitivity assumed in Refs.~\cite{Abbott:2017xzg,Aasi:2013wya}, though there are large uncertainties associated with both); current upper limits on the AGWB energy density are given by Refs.~\cite{TheLIGOScientific:2016dpb,TheLIGOScientific:2016xzw,LIGOScientific:2019vic,LIGOScientific:2019gaw}.
Once detected, the AGWB will give us a unique new probe of astrophysical processes and populations, spanning the entire history of cosmic star formation.
This extends the reach of BH and NS population studies to much higher redshift than is possible with individual CBC detections.

Much of the literature on the AGWB treats it, for simplicity, as perfectly isotropic.
However, recent work~\cite{Jenkins:2018lvb,Jenkins:2018uac,Cusin:2018rsq} has investigated how the anisotropic distribution of sources on the sky and the inhomogeneous geometry of the intervening spacetime induce anisotropies in the AGWB~\footnote{
    Note that since the AGWB is composed of finitely many sources that are localised in time, it will necessarily be anisotropic if observed for a finite amount of time, due to random fluctuations in the signal.
    This form of anisotropy has been investigated in Ref.~\cite{Meacher:2014aca}.
    However, the anisotropies that we study are due to inhomogeneities in the source distribution and the geometry of spacetime, meaning that they will not average to zero when integrating over long observation times.
    It is this form of anisotropy that can give us insights into LSS.}.
This invites the exciting prospect of using the AGWB to explore the large-scale structure (LSS) of the Universe, with the GW sources acting as tracers of the cosmic matter distribution.

In order to realise the full potential of the AGWB as a probe of LSS it is vital to understand the rates and distributions of CBCs, as these provide the link between the cosmic matter distribution and the GW signal we observe.
As demonstrated recently in Ref.~\cite{Wysocki:2018mpo}, it is possible to reconstruct this information in the form of parameterised phenomenological CBC distributions, using LIGO/Virgo detections to infer posterior probabilities for the distribution parameters.
In contrast, the two existing predictions for the AGWB anisotropies in Refs.~\cite{Jenkins:2018uac,Cusin:2018rsq} each assumed a particular CBC distribution, neglecting the broad range of distributions that are compatible with current data.
It is unclear \emph{a priori} how these astrophysical uncertainties will affect the AGWB predictions.
Indeed, there is a significant discrepancy between these two predictions, raising the question of whether or not this can be explained by the differing astrophysical models.

In this Letter, we investigate the impact of the CBC distribution on the spectrum of AGWB anisotropies, as calculated in Ref.~\cite{Jenkins:2018uac}.
Using the methods presented in Ref.~\cite{Wysocki:2018mpo}, we sample more than 10,000 possible CBC distributions supported by LIGO/Virgo detections from the first two observing runs (O1 and O2), using the analysis of Ref.~\cite{Jenkins:2018uac} to translate these into confidence intervals on the angular spectrum of the AGWB energy density.
We look exclusively at binary black hole (BBH) mergers here, but the analysis can be very easily extended to include binary neutron star (BNS) and black-hole--neutron-star (BHNS) mergers once we have multiple detections of these.

%%%%%%%%%%%%%%%%%%%%%%%%%%%%%%%%%%%%%%%%%%%%%%%%%%%%%%%%%%%%%%%%%%%%%%%%%%%%%%%%%
\paragraph{Inferring population parameters.}$\!\!\!\!$---
The most important quantities describing each BBH are the masses $m_1,m_2$ and dimensionless spin vectors $\vb*\chi_1,\vb*\chi_2$ of each component BH.
These are the intrinsic parameters determining the GW waveform produced by each BBH.
In order to calculate the net AGWB signal produced by a large number of BBHs, one needs an astrophysical distribution for each of these parameters.
These distributions are not known from first principles, but Bayesian techniques can be used to infer them from GW observations.
One also needs to know the local (i.e.,~redshift zero) rate of BBH mergers, $R_\mathrm{BBH}^{(\mathrm{local})}$, which is inferred in a similar way.
(See Ref.~\cite{Abbott:2017vtc} for the most recent BBH rate estimate.)

Population inference from GW observations proceeds via Bayes' theorem as usual, inferring the parameters of a presumed source population model via the likelihood of the number and nature of the data observed; see Ref.~\cite{Wysocki:2018mpo} for an introduction.
The likelihood function captures survey selection effects and statistical measurement errors, while a parametric source population model attempts to use a small number of population ``hyperparameters'' to broadly encode key features that should be qualitatively produced in many binary black hole formation models, particularly those derived from stellar-origin black holes.
Specifically, following Refs.~\cite{Wysocki:2018mpo,Fishbach:2017zga} we adopt a truncated power-law BH mass distribution,
    \begin{equation}
    \label{eq:bh-mass-distribution}
        p\qty(m_1,m_2)\propto
        \begin{cases}
            \frac{m_1^{-\alpha_m}}{m_1-m_\mathrm{min}},&
            \begin{matrix}
                m_\mathrm{min}\le m_2\le m_1\le m_\mathrm{max}\\
                m_1+m_2\le M_\mathrm{max}
            \end{matrix}\\
            0,&\mathrm{otherwise}
        \end{cases}
    \end{equation}
    where $m_\mathrm{min}=5M_\odot$ and $M_\mathrm{max}=200M_\odot$ are fixed, while $\alpha_m$ and $m_\mathrm{max}$ are inferred from observed BBHs.
(In Ref.~\cite{Jenkins:2018uac} they were fixed as $\alpha_m=2.35$ and $m_\mathrm{max}=95M_\odot$.)
In this model, the two mass limits $m_\mathrm{max},m_\mathrm{min}$ encode some approximately known maximum~\cite{Heger:2002by,Belczynski:2016jno} and minimum~\cite{Ozel:2010su,Farr:2010tu} masses set by the physics of stellar evolution; the power law encodes expected scaling derived from the stellar initial mass function (IMF), formation processes, and the strong dependence of the merger delay time on orbital period.
The model can predict observations dominated by massive BHs or by low-mass BHs, depending on the choice of power law exponent $\alpha_m$ and cutoff $m_\mathrm{max}$.
Similar to the presumed stellar IMF, this simple power law ansatz is an adequate phenomenological characterisation of the most observationally critical features of the BH population: the relative abundance of low-mass and very massive progenitors, versus a putative maximum BH mass.

The BH spin magnitudes are modeled by a Beta distribution,
    \begin{equation}
        p\qty(\chi_i)\propto\chi_i^{\alpha_\chi-1}\qty(1-\chi_i)^{\beta_\chi-1},
    \end{equation}
    where $\chi_i\equiv\qty|\vb*\chi_i|$, and the two parameters $\alpha_\chi,\beta_\chi$ are inferred.
(In Ref.~\cite{Jenkins:2018uac} the spin distribution was flat, corresponding to $\alpha_\chi=\beta_\chi=1$.)
For simplicity, we assume that the BH spins are either aligned or antialigned with the orbital axis (with equal probability), and therefore neglect precessing spins.
While spin precession effects are important for making precision measurements of individual BBH waveforms, we expect them to have negligible effect on the total GW energy density, particularly when averaged across a large ensemble of events.

Within the context of this model for the mass and spin distributions, Ref.~\cite{Wysocki:2018mpo} deduced the relative joint likelihood of different sets of the hyperparameters $R_\mathrm{BBH}^{(\mathrm{local})}$, $\alpha_m$, $m_\mathrm{max}$, $\alpha_\chi$, $\beta_\chi$.
(This was done using a prior uniform in $\alpha_m$, $m_\mathrm{max}$ and log uniform in $R_\mathrm{BBH}^{(\mathrm{local})}$, $\alpha_\chi$, $\beta_\chi$, with ranges sufficiently broad that the prior limits do not affect the support of the posterior.)
In this work we draw $\order{10^4}$ samples from their hyperparameter distribution~\footnote{%
    The code is publicly available at~\url{https://gitlab.com/dwysocki/bayesian-parametric-population-models}.
    Some examples of how to generate hyperparameter samples can be found at~\url{https://gitlab.com/dwysocki/pop-models-examples}, where the code is applied to the ten BBH detections catalogued in LIGO/Virgo GWTC-1~\cite{LIGOScientific:2018mvr}.
    }.%

%%%%%%%%%%%%%%%%%%%%%%%%%%%%%%%%%%%%%%%%%%%%%%%%%%%%%%%%%%%%%%%%%%%%%%%%%%%%%%%%%
\paragraph{The AGWB angular power spectrum.}$\!\!\!\!$---
\begin{figure*}[t]
    \includegraphics[width=0.495\textwidth]{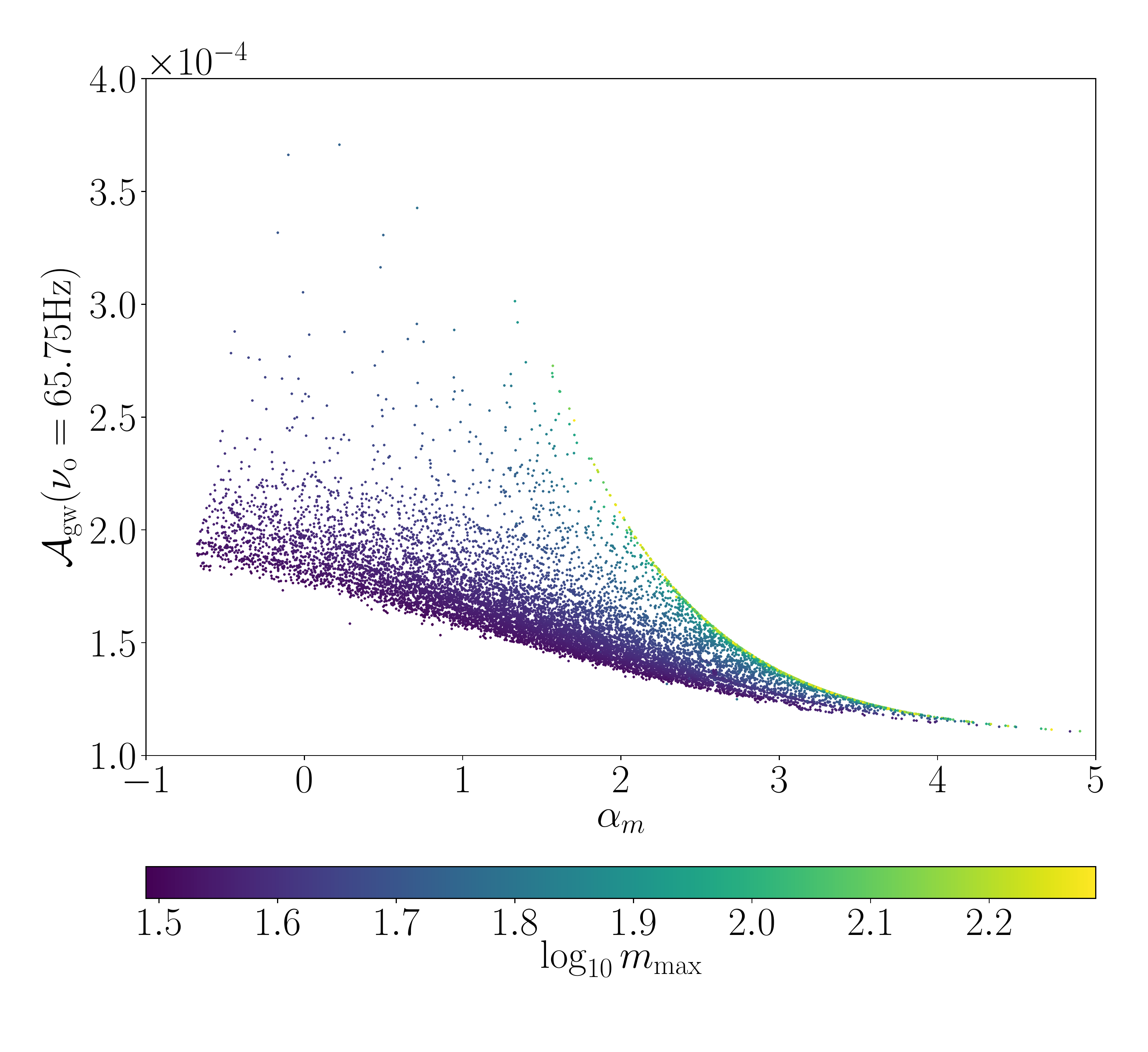}
    \includegraphics[width=0.495\textwidth]{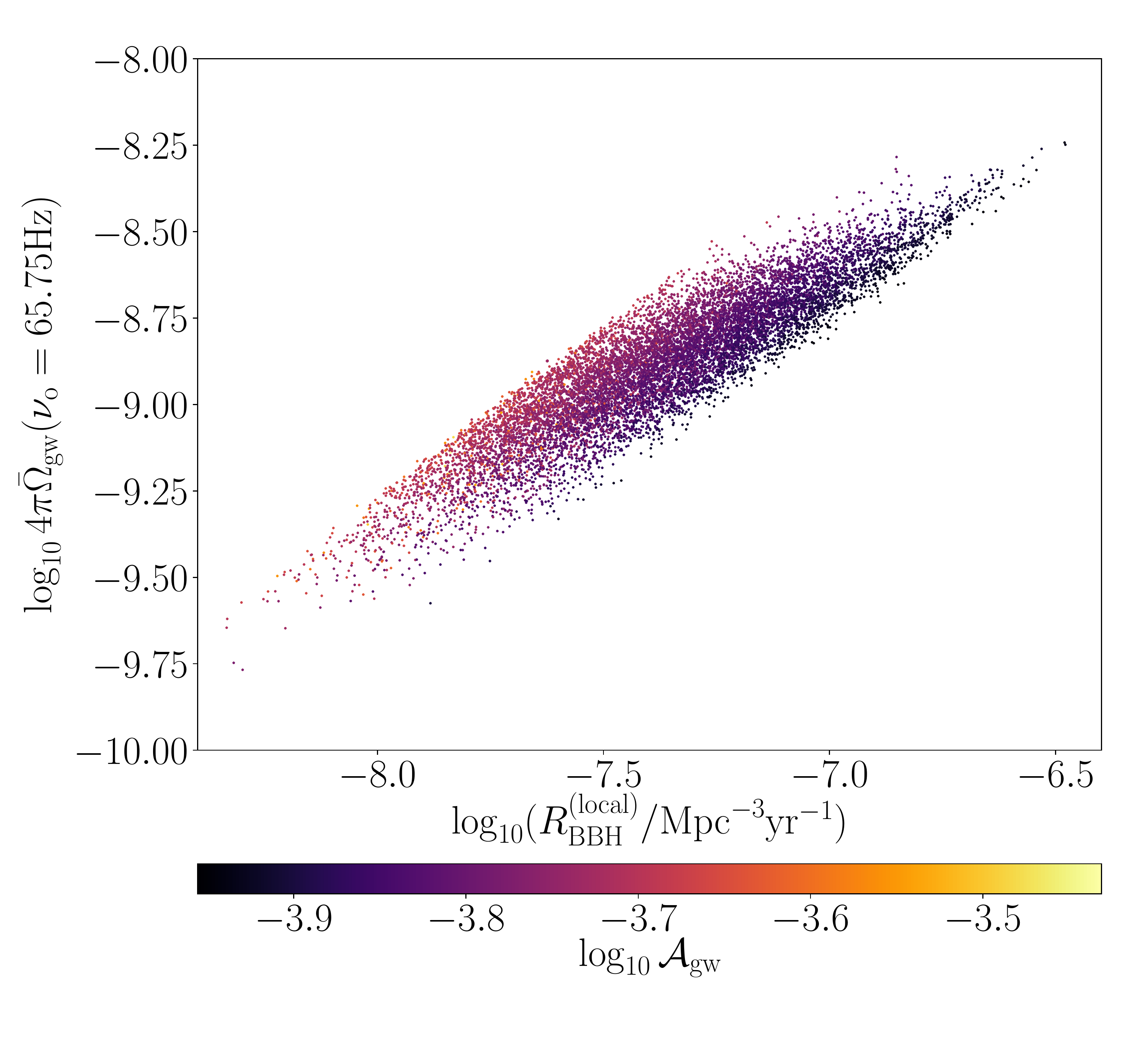}
    \caption{Left panel: AGWB anisotropy parameter $\mathcal{A}_\mathrm{gw}$ plotted against the BH mass power law exponent $\alpha_m$ for $\order{10^4}$ possible BBH distributions. The points are coloured according to the upper mass cutoff $m_\mathrm{max}$, which ranges between $30M_\odot$ and $200M_\odot$. Right panel: Total AGWB energy density $4\uppi\bar{\Omega}_\mathrm{gw}$ plotted against the local BBH rate $R_\mathrm{BBH}^{(\mathrm{local})}$ for $\order{10^4}$ possible BBH distributions. The points are coloured according to the anisotropy parameter $\mathcal{A}_\mathrm{gw}$.}
    \label{fig:A_gw-points}
\end{figure*}
We describe the AGWB in terms of its dimensionless density parameter,
    \begin{equation}
        \Omega_\mathrm{gw}\qty(\nu_\mathrm{o},\vu*e_\mathrm{o})\equiv\frac{1}{\rho_\mathrm{c}}\frac{\dd[3]{\rho_\mathrm{gw}}}{\dd{\qty(\ln\nu_\mathrm{o})}\dd[2]{\sigma_\mathrm{o}}},
    \end{equation}
    which represents the GW energy density per logarithmic frequency bin per solid angle $\dd[2]{\sigma_\mathrm{o}}$ in units of the cosmological critical density $\rho_\mathrm{c}$, where $\nu_\mathrm{o}$ is the observer-frame frequency and $\vu*e_\mathrm{o}$ is the observation direction.
The isotropic average (monopole) of the AGWB density parameter is
    \begin{equation}
        \label{eq:Omega-bar}
        \bar{\Omega}_\mathrm{gw}\qty(\nu_\mathrm{o})\equiv\frac{1}{4\uppi}\int_{S^2}\dd[2]{\sigma_\mathrm{o}}\Omega_\mathrm{gw}\qty(\nu_\mathrm{o},\vu*e_\mathrm{o}),
    \end{equation}
    with the relative size of fluctuations around this value described by the GW density contrast,
    \begin{equation}
        \delta_\mathrm{gw}\qty(\nu_\mathrm{o},\vu*e_\mathrm{o})\equiv\frac{\Omega_\mathrm{gw}-\bar{\Omega}_\mathrm{gw}}{\bar{\Omega}_\mathrm{gw}}.
    \end{equation}
This is a random field on the sphere, conveniently characterised by its two-point correlation function (2PCF),
    \begin{equation}
        C_\mathrm{gw}\qty(\nu_\mathrm{o},\theta_\mathrm{o})\equiv\ev{\delta_\mathrm{gw}\qty(\nu_\mathrm{o},\vu*e_\mathrm{o})\delta_\mathrm{gw}\qty(\nu_\mathrm{o},\vu*e_\mathrm{o}')},
    \end{equation}
    where the angle brackets denote an averaging over all pairs of points $\vu*e_\mathrm{o},\vu*e_\mathrm{o}'$ separated by an angle $\theta_\mathrm{o}=\cos^{-1}\qty(\vu*e_\mathrm{o}\vdot\vu*e_\mathrm{o}')$.
It is common practice to perform a multipole expansion of the 2PCF,
    \begin{equation}
        C_\mathrm{gw}\qty(\nu_\mathrm{o},\theta_\mathrm{o})=\sum_{\ell=0}^\infty\frac{2\ell+1}{4\uppi}C_\ell\qty(\nu_\mathrm{o})P_\ell\qty(\cos\theta_\mathrm{o}),
    \end{equation}
    decomposing in terms of the Legendre polynomials $P_\ell\qty(x)$.
The statistics of the AGWB anisotropies are then described by the $C_\ell$ components,
    \begin{equation}
        \label{eq:analytical-C_ell}
        C_\ell\qty(\nu_\mathrm{o})\equiv2\uppi\int_{-1}^{+1}\dd{\qty(\cos\theta_\mathrm{o})}C_\mathrm{gw}\qty(\nu_\mathrm{o},\theta_\mathrm{o})P_\ell\qty(\cos\theta_\mathrm{o}),
    \end{equation}
    which, roughly speaking, represent the magnitude of AGWB fluctuations on angular scales of $\uppi/\ell$.

Assuming that the BHs we observe with LIGO and Virgo are the result of stellar evolution (rather than being primordial in origin), they must reside in galaxies.
Thus in order to calculate the $C_\ell$'s, one needs to model the inhomogeneous distribution of galaxies.
In Ref.~\cite{Jenkins:2018uac} two different approaches are adopted to achieve this: (i) an approximate analytical model, where the average number density and anisotropic clustering of the galaxies are described by simple functional fits to galaxy survey data~\cite{2016ApJ...830...83C,Marulli:2013wpa}; (ii) a mock galaxy catalogue based on the Millennium simulation~\cite{Blaizot:2003av,DeLucia:2006szx,Springel:2005nw,Lemson:2006ee}.
In the analytical approach, one finds the simple expression
    \begin{equation}
    \label{eq:C_ell-analytical}
        C_\ell\qty(\nu_\mathrm{o})=4\uppi\mathcal{A}_\mathrm{gw}\qty(\nu_\mathrm{o})\frac{{}_3F_2\qty(-\ell,\ell+1,1-\frac{\gamma}{2};1,2;1)}{\mathrm{sinc}\qty(\uppi\gamma/2)},
    \end{equation}
    where $\mathcal{A}_\mathrm{gw}$ is a frequency-dependent co\"efficient that depends on the astrophysical model, ${}_3F_2$ is a generalised hypergeometric function, $\gamma$ is the slope of the power law describing the galaxy-galaxy 2PCF, and $\mathrm{sinc}\qty(x)\equiv\sin\qty(x)/x$.
In the catalogue approach, one sums the contributions from every single galaxy in the catalogue, weighted according to their GW flux, and uses HEALP\textsc{ix}~\footnotetext[2]{\url{http://healpix.sourceforge.net}}\cite{Note2,Gorski:2004by} to calculate the $C_\ell$'s from the resulting AGWB map.
Despite the simplicity of the analytical approximation, these two approaches are in excellent agreement on large angular scales, and differ only by an $\order{1}$ factor at small scales~\cite{Jenkins:2018uac}.

There are thus two ways that we can calculate the changes in the $C_\ell$'s resulting from different astrophysical distributions, following each of the two methods described above.

%%%%%%%%%%%%%%%%%%%%%%%%%%%%%%%%%%%%%%%%%%%%%%%%%%%%%%%%%%%%%%%%%%%%%%%%%%%%%%%%%
\paragraph{Results.}$\!\!\!\!$---
Using the results of Ref.~\cite{Wysocki:2018mpo}, we sample $\order{10^4}$ possible sets of BBH population hyperparameters, based on the LIGO/Virgo BBH detections from O1 and O2.
For each of these, we use the analytical method of Ref.~\cite{Jenkins:2018uac} to calculate the SGWB monopole $\bar{\Omega}_\mathrm{gw}$ and anisotropy parameter $\mathcal{A}_\mathrm{gw}$ [defined in Eqs.~\eqref{eq:Omega-bar} and~\eqref{eq:analytical-C_ell}, respectively---these can be evaluated at any GW frequency, but we choose $\nu_\mathrm{o}=65.75~\mathrm{Hz}$ as this is forecast to be the frequency of maximum sensitivity of the LIGO detector network to the AGWB~\cite{Abbott:2017xzg}].
This analysis shows that the anisotropies are largely insensitive to the details of the BBH population, with the value of $\mathcal{A}_\mathrm{gw}$ varying by just an $\order{1}$ factor between all the BBH distributions (in fact, more than 99\% of the $\mathcal{A}_\mathrm{gw}$ values are within a factor $2$ of each other).
Recall that in our convention for the $C_\ell$'s, it is only the \emph{relative} size of the anisotropies compared to the total energy density $4\uppi\bar{\Omega}_\mathrm{gw}$ that matters.
We find that the total energy density varies over nearly 2 orders of magnitude for the range of astrophysical distributions considered here (cf. the right panel of Fig~\ref{fig:A_gw-points}).

We can also use these results to explore whether the AGWB monopole and anisotropies are correlated with any of the population parameters used in our model.
In particular, there is an interesting relationship between the size of the anisotropies (given by $\mathcal{A}_\mathrm{gw}$), the BH mass power-law index $\alpha_m$, and the maximum BH mass $m_\mathrm{max}$, as illustrated in the left panel of Fig.~\ref{fig:A_gw-points}.
Increasing $m_\mathrm{max}$ leads to larger anisotropies, as this allows for more massive BBHs that contribute more strongly to the AGWB and therefore lead to larger energy density fluctuations.
This effect is suppressed as $\alpha_m$ is increased, as this causes the BBH population to become increasingly dominated by low-mass BHs.

There is also a relationship between $\mathcal{A}_\mathrm{gw}$, the monopole $\bar{\Omega}_\mathrm{gw}$, and the local BBH rate $R_\mathrm{BBH}^{(\mathrm{local})}$, as illustrated in the right panel of Fig.~\ref{fig:A_gw-points}.
Increasing the BBH rate generally increases the AGWB energy density due to the larger number of events.
However, for a fixed $\bar{\Omega}_\mathrm{gw}$ there is a spread of points, showing a trade-off between a higher rate of fainter BBHs or a lower rate of louder BBHs.
The latter leads to an AGWB that is more granular and dominated by large fluctuations, increasing the anisotropy parameter $\mathcal{A}_\mathrm{gw}$.

We also find that there is no correlation of the spin hyperparameters $\alpha_\chi,\beta_\chi$ with either $\mathcal{A}_\mathrm{gw}$ or $\bar{\Omega}_\mathrm{gw}$.
This shows that BH spins have negligible impact on the AGWB, which is intuitively reasonable, as the total energy radiated from a CBC is set primarily by the masses.

\begin{figure}[t]
    \includegraphics[width=0.5\textwidth]{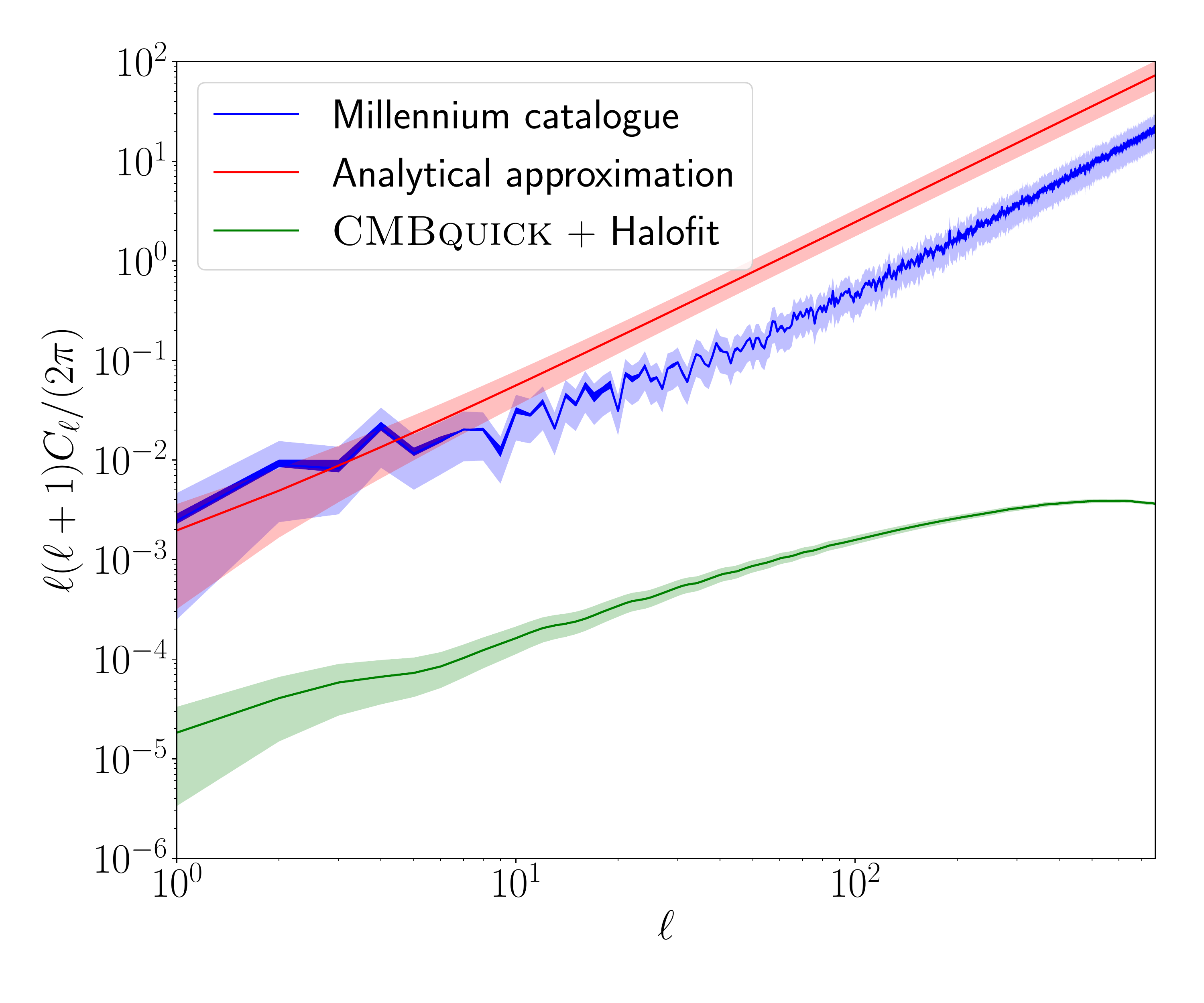}
    \caption{Various predictions for the $C_\ell$ spectrum of the AGWB overdensity field $\delta_\mathrm{gw}$, plotted as $\ell\qty(\ell+1)C_\ell/\qty(2\uppi)$, which is approximately the contribution to the overdensity field variance per logarithmic bin in $\ell$.
    The blue curve is calculated using the Millennium catalogue as in Ref.~\cite{Jenkins:2018uac}, with the dark blue region indicating the $99\%$ confidence interval when accounting for the BBH population uncertainties (based on $\sim$3000 random samples of the population hyperparameters), while the pale blue region indicates the $1\sigma$ error due to population uncertainty, cosmic variance, and finite-number Poisson uncertainty in the catalogue.
    The red curve is calculated using Eq.~\eqref{eq:C_ell-analytical}, with $\mathcal{A}_\mathrm{gw}=1.110\times10^{-4}$ corresponding to the maximum-likelihood BBH population parameters, and with the galaxy-galaxy two-point correlation power law index and clustering length set to $\gamma=1.67\pm0.03$ and $d_1=(5.05\pm0.26)h^{-1}~\mathrm{Mpc}$, respectively, to match the Millennium catalogue.
    The pale red region indicates the $1\sigma$ error due to uncertainty in $\mathcal{A}_\mathrm{gw}$, $\gamma$, and $d_1$, as well as cosmic variance.
    The green curve is calculated using \textsc{CMBquick}~\cite{Note3} and \textsc{Halofit}~\cite{Smith:2002dz,Note4} as in Ref.~\cite{Cusin:2018rsq}, using the maximum-likelihood BBH population, with the pale green region showing the $1\sigma$ error due to cosmic variance.}
    \label{fig:C_ell-astro-sigma}
\end{figure}

In addition to our results for $\mathcal{A}_\mathrm{gw}$, we use the catalogue method of Ref.~\cite{Jenkins:2018uac} to calculate the $C_\ell$'s for a range of BBH populations.
This process is much more computationally demanding than the analytical method, so we randomly select a subset of $\sim$3000 possible populations from our previous $\sim$10000 samples.
The resulting spread of values allows us to calculate confidence intervals for each $C_\ell$ due to population uncertainties, as shown in Fig.~\ref{fig:C_ell-astro-sigma}.
We find that the population uncertainties cause fluctuations in the $C_\ell$'s of typically $\sim3\%$, which is much smaller than the uncertainties due to cosmic variance and Poisson fluctuations in the catalogue.

An immediate consequence of this is that the discrepancy between the $C_\ell$ predictions of \citet{Jenkins:2018uac} and \citet{Cusin:2018rsq} cannot be explained by population uncertainties, and must instead be due to their differing treatments of the galaxy distribution.
We confirm this by using the methods of both of these papers to calculate the $C_\ell$'s, but with the same (maximum-likelihood) BBH distribution.
The predictions based on Ref.~\cite{Jenkins:2018uac} use the Millennium catalogue as described above, while those based on Ref.~\cite{Cusin:2018rsq} use \textsc{CMBquick}~\footnotetext[3]{\url{http://www2.iap.fr/users/pitrou/cmbquick.htm}}\cite{Note3} and \textsc{Halofit}~\footnotetext[4]{\url{http://www.roe.ac.uk/~jap/haloes/}}\cite{Smith:2002dz,Note4}.
For the latter, we input the same maximum-likelihood BBH population as for the former, and set all other details of the astrophysical model (the GW waveforms, the prescription for how metallicity affects the BBH rate, etc.) to be the same as in Ref.~\cite{Jenkins:2018uac}.
This is done by calculating the appropriate values for the functions $\mathcal{A}\qty(z,\nu_\mathrm{o})$ and $\mathcal{B}\qty(z,\nu_\mathrm{o})$, as defined in Eqs.~(5) and~(6) of Ref.~\cite{Cusin:2018rsq}.
By ensuring that all these details are the same in both methods, we achieve a fair comparison between them, with the galaxy distribution being the only remaining difference.
The results of this comparison are shown in Fig.~\ref{fig:C_ell-astro-sigma}, where we see that there is a discrepancy of more than 2 orders of magnitude, far larger than the errors due to uncertainty in the BBH population.
It seems increasingly likely that this discrepancy is due to the linear transfer functions and linear scale-invariant bias adopted in Ref.~\cite{Cusin:2018rsq} to model the galaxy number inhomogeneities, which could potentially fail to capture the nonlinear clustering of the AGWB.

%%%%%%%%%%%%%%%%%%%%%%%%%%%%%%%%%%%%%%%%%%%%%%%%%%%%%%%%%%%%%%%%%%%%%%%%%%%%%%%%%
\paragraph{Conclusion.}$\!\!\!\!$---
We have used BBH population inference to explore the impact of population uncertainties on the AGWB.
Our results show that the isotropic AGWB monopole $\bar{\Omega}_\mathrm{gw}$ is sensitive to the nature of the BBH population (particularly the local rate), while the anisotropic $C_\ell$ spectrum is only modified to within a few percent, at a level which is insignificant compared to other sources of uncertainty (such as cosmic variance).

The calculations performed in this work adopt a very simple redshift-independent population model, with only the merger rate varying with redshift.
This is justified by sparse observational data, a wide variety of possible BBH formation scenarios, and theoretical modelling challenges associated with each.
Even neglecting the question of redshift dependence, we do not consider the full range of astrophysically motivated models; see in particular Ref.~\cite{Talbot:2018cva}.
However, there is no reason to expect that our main findings (that the monopole is strongly dependent on the BH mass distribution, but the $C_\ell$'s are not) are peculiar to the population models considered here.
On physical grounds we expect that the $C_\ell$'s will only vary significantly in models with a strong sensitivity to the properties of the host galaxy.
In particular, highly metallicity-dependent formation scenarios could increase the sensitivity of the AGWB to the distribution of galaxy metallicities, possibly leading to larger anisotropies.
This will be investigated in future work.
Nonetheless, the results presented here demonstrate the robustness of the AGWB $C_\ell$ spectrum over a very broad range of source populations, highlighting its potential as a powerful new probe of LSS, and of late-Universe cosmology in general.

%%%%%%%%%%%%%%%%%%%%%%%%%%%%%%%%%%%%%%%%%%%%%%%%%%%%%%%%%%%%%%%%%%%%%%%%%%%%%%%%%
\begin{acknowledgments}
We thank Andrew Matas for his thorough and valuable comments.
This Letter has been assigned the document number LIGO-P1800320.
The Millennium Simulation databases used in this Letter and the web application providing online access to them were constructed as part of the activities of the German Astrophysical Virtual Observatory.
Some of the results in this Letter have been derived using the HEALP\textsc{ix} package~\cite{Gorski:2004by}.
A.C.J. is supported by King's College London through a Graduate Teaching Scholarship.
M.S. is supported in part by the Science and Technology Facility Council (STFC), United Kingdom, under the Research Grant ST/P000258/1.
R.O.S. and D.W. gratefully acknowledge NSF Grant No. PHY-1707965.
D.W. also acknowledges support from RIT through the CGWA SIRA initiative.
\end{acknowledgments}

%%%%%%%%%%%%%%%%%%%%%%%%%%%%%%%%%%%%%%%%%%%%%%%%%%%%%%%%%%%%%%%%%%%%%%%%%%%%%%%%%
\paragraph{Note added.}$\!\!\!\!$---
Soon after the first version of this Letter was posted online, a Comment~\cite{Cusin:2018ump} appeared from the authors of Ref.~\cite{Cusin:2018rsq}, criticising the accuracy of the analytical $C_\ell$ spectrum Eq.~\eqref{eq:C_ell-analytical}.
We have responded to this Comment in Ref.~\cite{Jenkins:2019cau}, emphasising that the analytical approach is only a simple approximation, and not a firm prediction.
While Eq.~\eqref{eq:C_ell-analytical} is useful for rapidly investigating and developing intuition for thousands of populations models, our main predictions are based on the Millennium catalogue, for which all the criticisms in Ref.~\cite{Cusin:2018ump} are irrelevant.
Furthermore, we argue in Ref.~\cite{Jenkins:2019cau} that the catalogue approach should give more accurate predictions than the linearised approach in Ref.~\cite{Cusin:2018rsq}, due to the importance of strongly nonlinear scales in calculating the $C_\ell$ spectrum.

%%%%%%%%%%%%%%%%%%%%%%%%%%%%%%%%%%%%%%%%%%%%%%%%%%%%%%%%%%%%%%%%%%%%%%%%%%%%%%%%%
\bibliography{sgwb-bbh-pops}
\end{document}